# Near-room-temperature ferromagnetic ordering in the pressure-induced collapsed-tetragonal phase in SrCo$_2$P$_2$

S. Huyan[1,2*], J. Schmidt[1,2*], A. Valadkhani[3*], H. Wang[4], Z. Li[1,2], A. Sapkota[1,2], J. L. Petri[5], T. J. Slade[1,2], R. A. Ribeiro[1,2], W. Bi[5], W. Xie[4], I. I. Mazin[6], R. Valenti[3†], S. L. Bud'ko[1,2†], and P. C. Canfield[1,2†]

[1] *Ames National Laboratory, US DOE, Iowa State University, Ames, Iowa 50011, USA*
[2] *Department of Physics and Astronomy, Iowa State University, Ames, Iowa 50011, USA*
[3] *Institut för Theoretische Physik, Goethe-Universität, 60438 Frankfurt am Main, Germany*
[4] *Department of Chemistry, Michigan State University, East Lansing, MI 48824, USA*
[5] *Department of Physics, University of Alabama at Birmingham, Birmingham, AL 35294, USA*
[6] *Department of Physics and Astronomy, and Quantum Science and Engineering Center, George Mason University, Fairfax, VA, USA*

[†]*e-mail address: valenti@itp.uni-frankfurt.de*
*budko@ameslab.gov*
*canfield@ameslab.gov*

*\* Equally contributed*

**Abstract:**

We present high pressure electrical transport, magnetization, and single crystal X-ray diffraction data on SrCo$_2$P$_2$ single crystals. X-ray diffraction data show that there is a transition to a collapsed tetragonal structure for $p \gtrsim 10$ GPa and measurements of resistance show that above ~ 10 GPa, a clear transition-like feature can be observed at temperatures up to 260 K. Further magnetization, magnetoresistance and Hall measurements made under pressure all indicate that this transition is to a ferromagnetic ground state. First principles-based density functional theory (DFT) calculations also show that there is a first-order transition between tetragonal and collapsed tetragonal (cT) phases, with an onset near ~ 10 GPa as well as the appearance of the ferromagnetic (FM) ordering in the cT phase. Above ~ 30 GPa, the experimental signatures of the magnetic ordering vanish in a first-order-like manner, consistent with the theoretical calculation results, indicating that SrCo$_2$P$_2$ is another example of the avoidance of quantum criticality in ferromagnetic intermetallic compounds. SrCo$_2$P$_2$ provides clear evidence that the structural, electronic and magnetic properties associated with the cT transition are strongly entangled and are not only qualitatively captured by our first principles-based calculations but are quantitatively reproduced as well.

**Introduction:**

AT$_2$X$_2$ and AET$_2$X$_2$ (A=alkali metal, AE=alkali-earth metal, T=transition metal, X=pnictogen) compounds have garnered renewed attention since the discovery of superconductivity in the Fe-based family AFe$_2$As$_2$ [1]. With the discovery of CaFe$_2$As$_2$ [2] followed, in turn, by the discovery of a pressure induced, isostructural phase transition associated with As-As bonding across the Ca-layer, [3-9] the term collapsed tetragonal (cT) phase was coined [10], and revived interest in the bonding ideas introduced by Hoffman et al., [11] was generated. Although the most critical factor for the formation of collapsed phase has been disputed [12], the cT transition is characterized by a significant decrease in the X-X distance and a commensurate large change in the *c*-lattice parameter as well as a relatively smaller expansion of the *a*-lattice parameter [13-15]. Given this large change in bonding as well as lattice parameters, it is not surprising that a cT transition is accompanied by significant changes in the electronic, magnetic and mechanical properties of the system. In addition, there is an inherent degree of very strong electron-phonon coupling for displacements along the *c*-axis, specifically associated with the modulation of the X-X distance [8].

For example, CaFe$_2$As$_2$, either doped with cobalt or with post growth, annealing-controlled internal strain, can change from having either a superconducting or antiferromagnetically ordered ground state in the uncollapsed phase to being non-moment bearing and non-superconducting in the cT phase [16-20]. In the case of SrCo$_2$As$_2$ the cT transition changes the ground state from non-magnetic to magnetically ordered [21] and in the case of LaRu$_2$P$_2$ there is a bulk, superconducting (SC) ground state in the uncollapsed tetragonal (ucT) structure that disappears with the advent of the cT state [22-24]. In addition to dramatic changes in electronic and magnetic ground states, cT transition physics has now been associated with remarkable, superelastic phenomena in CaFe$_2$As$_2$ [25], CaKFe$_4$As$_4$ [26], and SrNi$_2$P$_2$ [27], leading to a maximum recoverable strain values of ~ 17% and deformation-free cycling of over 10,000 cycles [25, 27].

Recently the Sr(Ni$_{1-x}$Co$_x$)$_2$P$_2$ series was studied [28]. Although neither end member, SrNi$_2$P$_2$ and SrCo$_2$P$_2$, manifest magnetic ordering down to lowest measured temperatures, it was found that small moment ordering can be stabilized for $0.65 < x < 0.99$. These data very clearly showed that SrCo$_2$P$_2$ was exceptionally close to manifesting a low temperature, magnetically ordered state. This conclusion is consistent with earlier studies of (Sr$_{1-x}$Ca$_x$)Co$_2$P$_2$ [29] and SrCo$_2$(P$_{1-x}$Ge$_x$)$_2$ [30] that also showed that SrCo$_2$P$_2$ could be tuned (by smaller substitutions) to show magnetism. All of these results, particularly considering the change of volume due to the substitution of Sr by Ca and P by Ge, indicate that hydrostatic pressure should be a promising tuning mechanism. Given that SrCo$_2$P$_2$ is in the ucT structure, the possibility of inducing a cT state that could be magnetically ordered was an option to explore. Here we combine pressure dependent structural, electronic, and magnetic measurements with band-structure calculations to determine the *T-p* phase diagram of SrCo$_2$P$_2$ (Figure 1), and show that for pressures near 10 GPa there is a clear ucT to cT transition that is accompanied by the appearance of a ferromagnetic state with Curie temperature ($T_C$) ~ 250 K. With increasing pressure $T_C$ first increases to ~ 260 K and then decreases to 160 K by ~ 27 GPa. For higher pressures, $T_C$ discontinuously drops to zero, consistent with the avoided quantum criticality in ferromagnetic intermetallic compounds [31-44]. First principles density functional theory (DFT) calculations capture the physics of both the pressure induced cT transition as well as the ferromagnetic phase, accurately, and quantitatively identifying the pressures for (i) the cT and simultaneous FM transitions as well as (ii) the subsequent offset of the ferromagnetic (FM) state. Our findings demonstrate that pure SrCo$_2$P$_2$ is remarkably close to near-room-temperature ferromagnetism (at least barn temperature for an Iowa winter). Perhaps more importantly, we demonstrate that DFT calculations can quantitatively predict the critical pressure needed to induce the cT transition and also predict the critical pressure associated with the discontinuous loss of FM order at higher pressures. The remarkable agreement between the experimental and theoretical results provides a promising example of the predictive power of ab initio-based DFT calculations regarding structural and magnetic phases/phase transitions.

**Results and discussion:**

Figure 1 presents a concise summary of our primary experimental and computational results. After reviewing these results, we will present and discuss the specific data sets used to create the *T-p* diagram shown. Fig. 1(a) plots (i) the structural *c/a* ratio, and (ii) the ferromagnetic Curie temperature, $T_C$ as a function of applied pressure. For *p* < ~ 10 GPa, the *c/a* ratio gradually decreases with pressure. Near ~ 10 GPa the *c/a* ratio has a sharp, discontinuous decrease in value and then, for *p* > ~ 10 GPa, resumes a more gradual decrease with increasing pressure. We have found *c/a* to be the most informative, single parameter to track for ucT-cT transition since both the *c*- and *a*-lattice change, often in different directions. (Note that diffraction data are shown in supporting information (SI) Figs. S2-S4.) Our DFT results for *c/a* are also presented in the same plot, and show excellent agreement with the experimental data. Fig. 1(b) displays the room temperature resistance as a function of pressure for samples 2 and 3 (s2, s3). They show a significant increase in resistance as pressure exceeds 10 GPa, a result consistent with the discontinuous change in the lattice parameters. Returning to Fig. 1(a), for $p \lesssim 10$ GPa we do not detect any evidence of magnetic ordering. For $p \gtrsim 10$ GPa, we see experimental evidence of a FM state existing below a pressure-dependent $T_C$. Initially $T_C$ is relatively pressure insensitive and stays near 260 K, but then between 18 to 27 GPa, $T_C$ drops to ~ 160 K. For $p \gtrsim 28$ GPa $T_C$ discontinuously drops to below base temperature and once again SrCo$_2$P$_2$ appears to have a non-moment bearing ground state. Fig. 1(c) shows the pressure dependence of the ordered moment per cobalt atom in the ground state as determined by DFT. The ~10 GPa onset to a magnetically ordered state is clearly captured (as was the cT transition in Fig. 1(a)). In addition, the DFT calculations predict a discontinuous loss of ferromagnetism for $p \gtrsim 30$ GPa (depending on the model); this pressure is remarkably close to the experimentally measured $p \sim 28$ GPa disappearance of the experimental signatures of the FM transition.

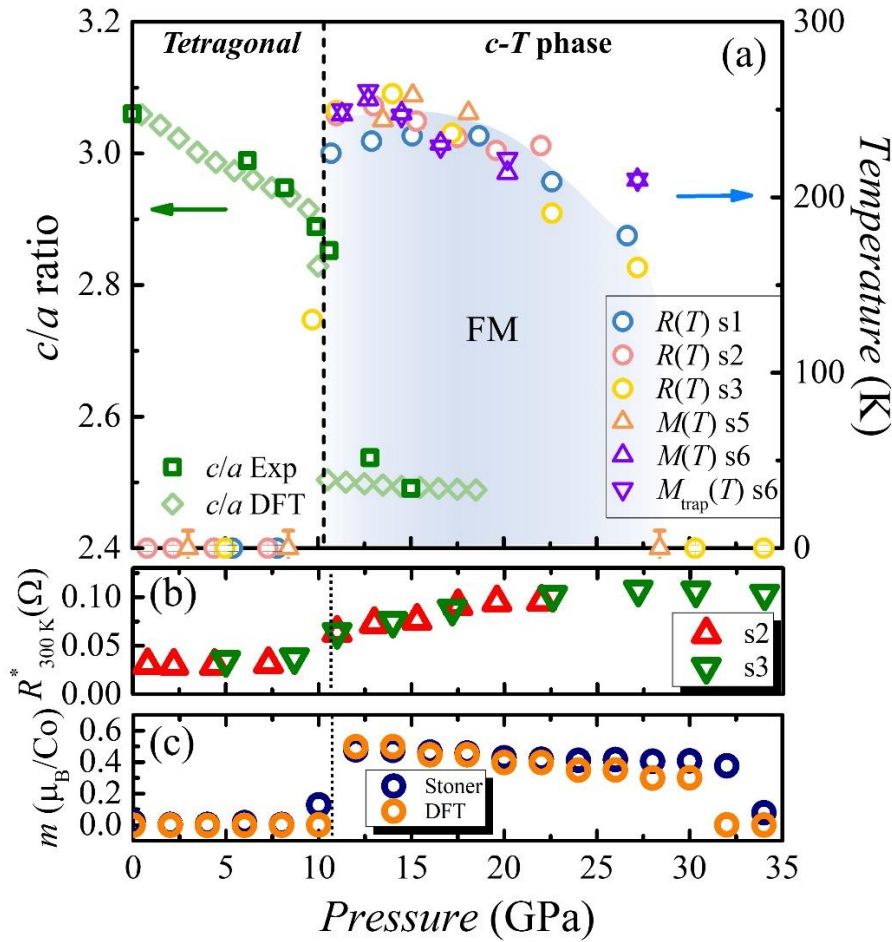

Fig. 1 **Pressure dependence of electronic, magnetic, and structural properties of SrCo$_2$P$_2$.** (a) Pressure dependent *c/a* ratio (left axis) determined from both experimental results and DFT calculations, and the ferromagnetic Curie temperature $T_C$ (right axis) inferred from electrical resistance and magnetization measurements on multiple samples denoted s1-s6. (b) Pressure dependent, 300 K resistance values of samples (s2 and s3). To allow comparison of the two data sets, the s2 data were normalized to s3 data at their shared 11 GPa $R(T)$ data point. Given this normalization, we denote the data as $R^*_{300K}$ in (b). The original plots for all studied samples are provided in Fig. S1 in the supporting information (SI). (c) The predicted ordered magnetic moment by both extended Stoner model and DFT calculations, as function of the pressure. The detailed description of theoretical calculations is shown in SI, Fig S12, S13.

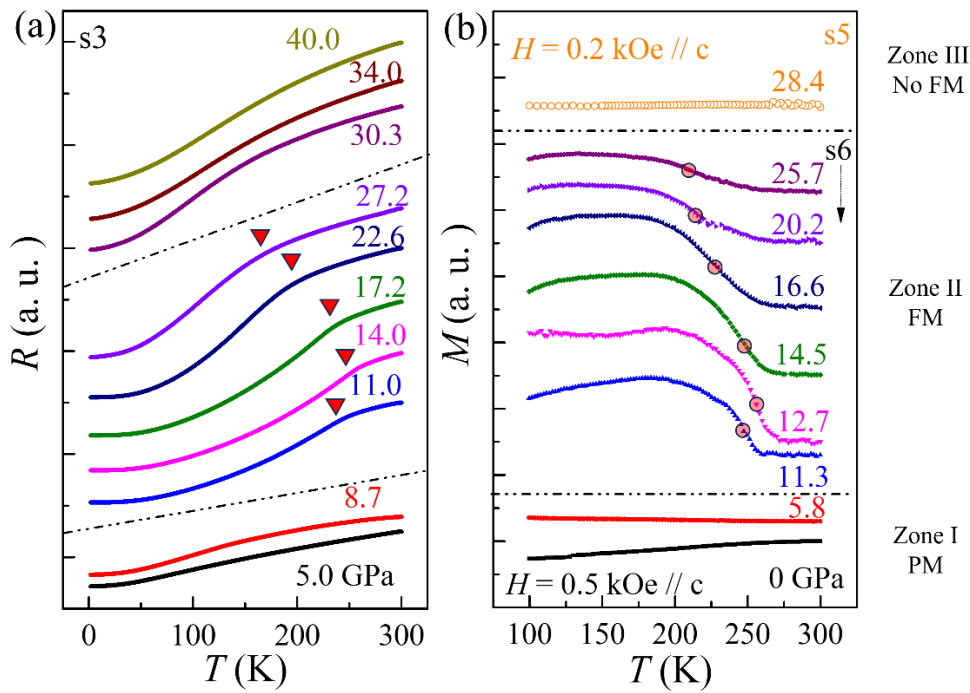

Fig. 2 Temperature dependence of (a) resistance data from s3, and (b) DC magnetization from s6 and s5 with 0.5 kOe and 0.2 kOe applied magnetic fields (respectively) applied parallel to the crystallographic *c*-axis, under high pressures on SrCo$_2$P$_2$ single crystals. All data are shifted along the y-axis for convenience of reading. Complete sets of data and the criteria used for determining transition temperatures are shown in Fig. S5, S6, S7, and S8 in the SI. The red triangles and pink circles mark the appearance, and the evolution of the phase transition observed in resistance and magnetization, respectively. We used the midpoint of the d$R$/d$T$ shoulder (shown in Fig. S6) and the extrema of the d$M$/d$T$ data (maximum negative slope, shown in Fig. S8 (b)), as the criteria of the $T_C$.

The emergence of (magnetic) ordering under pressure can be readily seen in electrical resistance measurements. Four samples were measured in diamond anvil cells (DACs) with different diamond culet sizes to cover different pressure ranges. The full data sets are plotted in Fig. S5. Fig. 2(a) shows the data from s3 and summarizes the overall picture of resistive behavior as a function of temperature of SrCo$_2$P$_2$ single crystal under pressure. At low pressure range, $p \lesssim 10$ GPa, sample shows metallic behavior from 1.8 K to 300 K without discernible anomalous features. Above ~10 GPa samples show significant increases in resistance, as well as a clear transition-like feature (slope change) with the ordering temperature, $T_C$, around 250 K at 11 GPa. The transition temperature $T_C$, which is defined as the midpoint of the shoulder of the d$R$/d$T$ peak, as shown in Fig. S6 in SI, forms a "dome" in $T_C$ versus pressure $p$ (Fig. 1(a)) by first increasing with $p$ to its maximum around 260 K at ~ 13 GPa, then $T_C$ being insensitive to the pressure till ~ 18 GPa, above which it decreases with pressure to ~160 K at 27 GPa, and then discontinuously drops to below base temperature at ~ 30 GPa. As discussed in the SI section, for pressure near 27 GPa determination of $T_C$ becomes harder, so the precise pressure at which $T_C$ drops to zero is unclear.

Magnetization measurements under pressure also show signatures of a ferromagnetic state being stabilized; a sharp increase in field cooled (FC) magnetization was observed at 11.3 GPa with the onset temperature ~ 247 K, consistent with the resistivity data. Here the criteria for defining $T_C$ are the minimum of the d$M$/d$T$ data (maximum negative slope), as shown in Fig. S8(b) in SI. The full FC data of s5 and s6 are shown in Fig. S7 and S8. At 13.5 GPa, the sample exhibits the highest magnetization below $T_C$, and further increase in pressure results in the suppression of ferromagnetic (FM) ordering, which becomes undetectable or barely discernible at approximately 28.4 GPa. To further examine the pressure stabilized FM state, thermoremanent magnetization measurements were also conducted on sample s6 (see Fig. S8(c)), where a noticeable change in slope was observed around ~ $T_C$, consistent with a ferromagnetic state with finite domain wall pinning.

It is noteworthy that during the ucT to cT transition, a distinct change in resistance is observed (Fig. 1(b) and Fig. S1). Although this could also be due to the changes in the carrier mobility [46], given that this $T > T_C$ increase in resistance essentially goes away for $T \ll T_C$ (as shown most clearly in figures S5(a and b), this anomaly is most likely associated with enhanced spin disorder scattering in the paramagnetic region. To further investigate the nature of the higher pressure phases, we performed transverse magnetoresistance measurements. The $R(T, H)$ data in the pressure region where the transition feature could be observed, e.g. 14 GPa, as presented in Fig. 3(a), has the resistive feature associated with the PM-FM transition broaden and shift to higher temperatures as the applied magnetic field is increased. This is consistent with the resistive signature of a FM transition in a metallic sample. Meanwhile, a negative magnetoresistance appears around the Curie temperature (Figs. 3(a) and 3(d)) owing to the spin-dependent scattering in magnetic systems. Such behavior was not observed at and above ~ 30 GPa (see Figs. 3(b) and 3(e)). Transverse magnetoresistance data taken for representative temperatures are shown in Figs. 3(c-e). At 14 GPa, a non-saturated negative magnetoresistance was observed which is consistent with the expectation for itinerant ferromagnets where the field dependence of the carrier fluctuation scattering can dominate, particularly near phase transitions [45]. Whereas, at lower pressure range, e.g. 5.3 GPa, and at higher pressure range, e.g. 30.3 GPa, magnetoresistance behaves as in a normal, non-magnetic, metal, as shown in Fig. 3(c) and 3(e), respectively. In addition, the results of Hall resistance are also consistent with FM order in $8.7 \lesssim p \lesssim 27.2$ GPa pressure range (see Fig. S9 in SI).

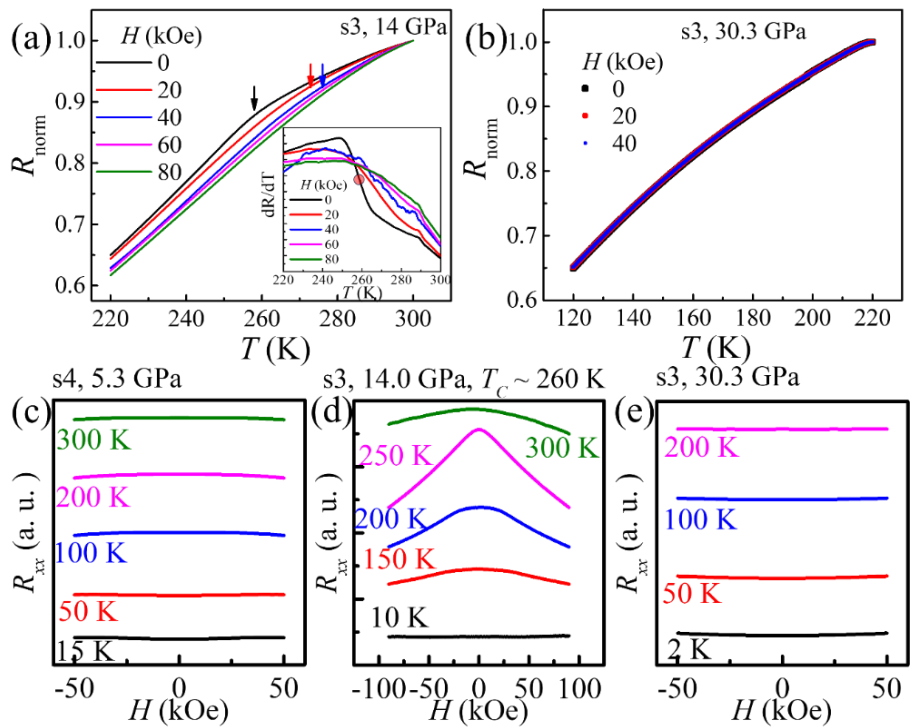

Fig. 3 **Magnetoresistance under pressure.** (a), (b) Temperature dependence of the normalized longitudinal resistance, $R_{norm}(T) = R(T) / R_{300K}$, at various magnetic fields, for 14 GPa and 30.3 GPa, respectively (both on s3). Arrows in (a) mark break in slope associated with mid-point of $dR/dT$ shoulder shown in the inset (c), (d), (e) Field dependent longitudinal magnetoresistance, $R_{xx}(H)$, at various temperature, for 5.3 GPa (on s3), 14 GPa, and 30.3 GPa (on s4), respectively.

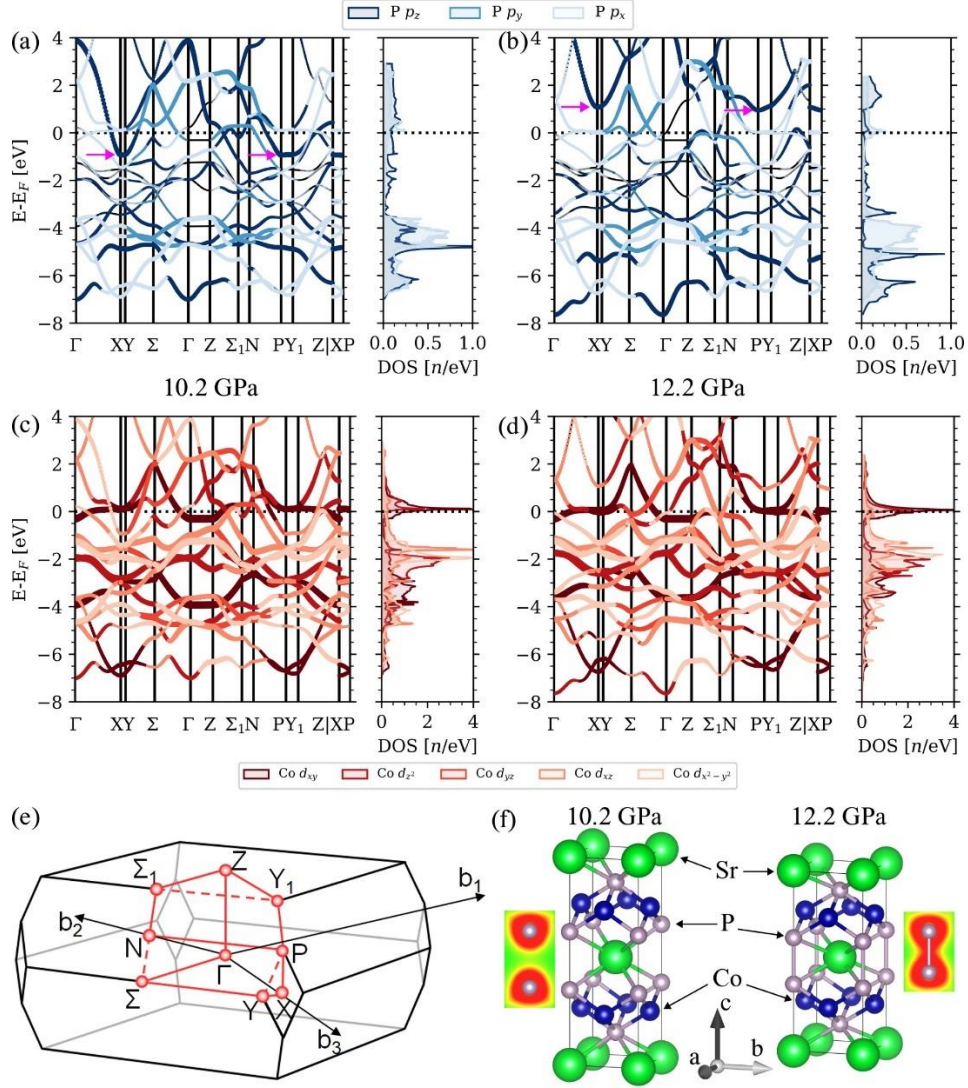

Fig. 4: (a) to (d) band structure of $SrCo_2P_2$ obtained within DFT in the GGA approximation. Magnetism is omitted in this calculation. (a) and (c) show the band structure in the ucT phase while (b) and (d) show the band structure in the cT phase. (a) and (b) focus on the P p-orbitals, which show a strong reconfiguration of the antibonding $p_z$ orbital (dark blue), as they become energetically unfavorable after the collapse (b) compared to before the collapse (a). (c) and (d) show the Co d-orbitals, where a van-Hove singularity near the Fermi level is observed showing as a very well defined sharp peak of the electronic density of state (DOS) near the Fermi level ($E_F$). As pressure increases beyond the transition, this peak broadens and eventually dissolves under higher pressures. (e) shows the Brillouin zone of the primitive unit cell. (f) shows the crystal structure in the conventional unit cell before, and after the collapse, with the charge density around the P atoms clearly showing bond formation in cT phase.

Band structure calculations, using DFT with (generalized gradient approximation) GGA, of $SrCo_2P_2$ are shown in Fig. 4. It should be noted that in order to correct for the common difference between DFT-based lattice parameters and those collected experimentally we simply calculated the pressure dependence of the c/a ratio and shifted it along the pressure scale to agree with the ambient pressure data. The details of the calibration strategy are provided in the methods section below as well as in Fig. S10. Post-calibration, we can quantitatively predict the pressure needed to induce the cT phase transition as well as subsequent magnetic transitions.

Focusing on the $p_z$ character of the P-bands, shown in Fig. 4(a) and (c), at 10.2 GPa, we can mainly see two electron pockets crossing the Fermi level at the *X, Y* and *P* points, representing the P-P $p_z$ anti-bonding bands. At 12.2 GPa, these anti-bonding bands move well above the Fermi level, indicating depopulation of the anti-bonding states, which suggests a stronger P-P bonding character of the collapsed phase (as shown in real-space in Fig. 4(f)).

For the $d_{xy}$ character of the Co bands, as shown in Fig. 4(b) and (d), there are relatively flat bands slightly above the Fermi level. Upon applying more pressure, these flatter bands start to cross the Fermi level, increasing the density of states (DOS) at the Fermi level. This very likely is associated with the emergence of itinerant ferromagnetism in the context of the Stoner criterion. A clear DOS peak is observed, indicating an instability towards uniform spin alignment, which leads to ferromagnetic order.

By calculating the total energy for fixed spin moments, we can determine the size of the ordered moment as a function of pressure by identifying the energetic minima (Fig. S11). The locus of the minima qualitatively reproduces the *T-p* phase diagram shown in Fig. 1, including the discontinuous loss of the FM phase for pressures greater than ~ 30 GPa. The DFT-calculated amplitudes of the ordered moments as a function of pressure are shown in Figs. S12 and S13.

We have been able to create a comprehensive framework of understanding of the pressure dependence of $SrCo_2P_2$ based on pressure and temperature dependent measurements of resistance, magnetoresistance, Hall resistance, magnetization, and crystal structure coupled with pressure dependent DFT-based calculations. $SrCo_2P_2$ has a pressure induced ucT-cT transition for $p \sim 10$ GPa with the cT phase manifesting a near-room-temperature FM ordering for $\sim 10$ GPa $< p < \sim 27$ GPa, above which the FM discontinuously disappears. $SrCo_2P_2$ demonstrates that (i) small moment magnetism in the cT phase can be induced by modest pressures, (ii) the FM state in the cT phase provides another example of avoided FM QCP, and (iii) DFT analysis can capture virtually all of the salient physics associated with the cT transition, quantitatively predicting critical pressures for multiple phase transitions.

From a potential applications standpoint, it is noteworthy that the transition from the tetragonal phase to the collapsed tetragonal phase is characterized by a significant reduction in the lattice parameter *c* and an expansion in the lattice parameter *a*. As has been also demonstrated for other 122 materials [25-27], strain along *c*, rather than hydrostatic pressure, can be an effective method for tuning the structure and magnetic ordering in the case of $SrCo_2P_2$. Whereas the ~ 10 GPa needed to induce the FM cT state in $SrCo_2P_2$ is rather high, the fact that DFT can quantitatively predict the pressure needed to induce the cT phase and also evaluate the size of the ordered moment means that new examples of ucT-cT transitions with tunable magnetic ordering can be found with the clear guidance of high-throughput computational studies of promising structures and compositions. Ultimately, the strong coupling between pressure/strain, structure, band-structure and magnetism offers the possibility of multiparameter tuning of these versatile systems. As such, the remarkable agreement between the pressure dependent experimental and computational results for $SrCo_2P_2$, specifically quantitative predictions of critical pressures for multiple phase transitions, provides a particularly clear example of the predictive power of ab initio-based DFT calculations for control complex and promising intermetallic systems.

**Acknowledgements:**

Work at Ames National Laboratory is supported by the US DOEf, Basic Energy Sciences, Material Science and Engineering Division under contract no. DE-AC02-07CH11358. I.I.M. acknowledges support from the National Science Foundation under Award No. DMR-2403804, and is also thankful to the Wilhelm and Else Heraeus Foundation for supporting his visits to the University of Frankfurt. R.V. and A.V. gratefully acknowledge support by the Deutsche Forschungsgemeinschaft (DFG, German Research Foundation) for funding through TRR 288 - 422213477 (project A05). W. X. and H-Z W are supported by the U. S. Department of Energy (DOE), Office of Science, Basic Energy Sciences under award DE-SC0023648. W.B. acknowledges the support from National Science Foundation (NSF) CAREER Award No. DMR-2045760. COMPRES-GSECARS gas loading system, and the $PX^2$ program were supported by COMPRES under NSF Cooperative Agreement No. EAR-1606856. We thank Dongzhou Zhang for assistance in the high-pressure XRD experiments.

# Supporting information

## Near-room-temperature ferromagnetic ordering in the pressure-induced collapsed-tetragonal phase in SrCo$_2$P$_2$

S. Huyan[1,2*], J. Schmidt[1,2*], A. Valadkhani[3*], H. Wang[4], Z. Li[1,2], A. Sapkota[1,2], J. L. Petri[5], T. J. Slade[1,2], R. A. Ribeiro[1,2], W. Bi[5], W. Xie[4], I. I. Mazin[6], R. Valenti[3†], S. L. Bud'ko[1,2†], and P. C. Canfield[1,2†]

[1] Ames National Laboratory, US DOE, Iowa State University, Ames, Iowa 50011, USA
[2] Department of Physics and Astronomy, Iowa State University, Ames, Iowa 50011, USA
[3] Institut für Theoretische Physik, Goethe-Universität, 60438 Frankfurt am Main, Germany
[4] Department of Chemistry, Michigan State University, East Lansing, MI 48824, USA
[5] Department of Physics, University of Alabama at Birmingham, Birmingham, AL 35294, USA
[6] Department of Physics and Astronomy, and Quantum Science and Engineering Center, George Mason University, Fairfax, VA, USA

**Methods:**

*Sample preparation*

Single crystals of SrCo$_2$P$_2$ were synthesized using a high-temperature solution growth method in Sn flux [1,2]. High purity elements were loaded into a 2 ml alumina fritted Canfield crucible set [3,4] and sealed under a partial argon atmosphere in a fused silica tube. The initial composition for growth was Sr$_{1.2}$Co$_2$P$_2$Sn$_{20}$. The ampoule was placed in a box furnace and held at 600°C for 6 hours, then heated to 1180°C and maintained at that temperature for 24 hours to ensure complete melting of the materials. Single crystals were grown by slowly cooling the ampoule over 100 hours to 950°C, at which point the excess Sn was decanted using a centrifuge. Powder X-ray diffraction measurement was performed using Rigaku-Miniflex Powder X-ray Diffractometer to confirm the phase.

*Electrical transport*

The electrical resistance measurements with current applied within the ab plane and magnetic field parallel to c-axis were performed in a Quantum Design Physical Property Measurement System (PPMS). The crystal was cleaved and cut into a thin regular shape for the high pressure electrical measurements so that slight Sn surface contamination could be mostly avoided. A standard, linear four-probe method (samples s1 - s4), and an approximately standard four terminal configuration for Hall measurements (sample s6) were used in a DAC. We applied 10 kOe and - 10 kOe magnetic field to get the magnetoresistance by ($R$ (10 kOe, $T$) + $R$ (-10 kOe, $T$)) / 2, and Hall resistance by ($R$ (10 kOe, $T$) – $R$ (-10 kOe, $T$)) / 2, respectively. For different pressure ranges, a diamond anvil cell (DAC) (Bjscistar, [5]) that fits into a Quantum Design Physical Property Measurement System (PPMS) was used with 500 μm (samples s1, s2 up to ~ 25 GPa), 400 μm (samples s3, s6 up to 40 GPa), and 700 μm (sample s4, up to ~ 6 GPa) culet-size standard-cut type Ia diamonds. The samples were cut and polished into rectangular flakes, with all samples around 100×40×15 μm in size, and loaded together with small ruby spheres into apertured stainless-steel gaskets, covered by cubic boron nitride (c-BN). The hole size of the gasket was roughly one third the diameter of diamond's culet. Platinum foil was used as the electrodes to connect to the sample. Nujol mineral oil was used as pressure transmitting medium (PTM), since: 1) fluid medium could still maintain a quasi-hydrostatic pressure environment with small pressure gradient below a liquid/glass transition which occurs at ~ 15 GPa [6-9]; 2) the use of fluid medium avoids the direct contact between the sample and diamond culet which will lead to a further contribution of uniaxial pressure component. Pressure was determined by the $R$1 line of the ruby fluorescent spectra [10,11].

*Magnetization*

The DC magnetization measurements under high pressure were performed in a Quantum Design Magnetic Property Measurement System (MPMS) at temperatures down to 5 K. The DAC (easyLab® Mcell Ultra [12]) with a pair of 500-μm-diameter culet-sized diamond anvils was used. The apertured tungsten gasket with 300-μm-diameter hole was used to lock the pressure. Nujol mineral oil, same as DAC for electrical transport measurement, was used to keep the consistency of pressure environments in two sets of experiments. The applied pressure was measured by the $R$1 fluorescence line of the ruby sphere [10,11]. The background signal of DAC without a sample was measured under ambient pressure. This included field-cooled (FC) magnetization measurements $M(T)$ under applied fields of 0.2, 0.5, and 1 kOe. Following this, the DAC was opened and re-sealed after loading the samples and ruby spheres, with dimensions of 200 μm × 200 μm × 20 μm (s5 for run 1) and 220 μm × 220 μm × 30 μm (s6 for run 2). The same as previous background measurements were then performed at various pressures. The magnetization of the sample was analyzed by first performing a point by point subtraction of long-scan response with/without the sample, and then a dipole fitting (a general fitting algorithm used to convert the raw SQUID voltage as a function of position, during standard DC scan, to the sample moment, using the Quantum Design MPMS.) of the subtracted long-scan response curve [13]. Additionally, a thermoremanent magnetization measurement was performed by cooling the DAC from 300 K to 70 K in a 20 kOe external field, stabilizing the temperature at 70 K for 10 minutes, and then removing the field. The warming up $M_{Trap}(T)$ measurement was carried out from 70 K to 300 K.

*X-ray diffraction*

To investigate the pressure effects on the crystal structure, high pressure single crystal X-ray diffraction (XRD) experiments were performed on a single crystal of $SrCo_2P_2$ with the dimensions of $0.088 \times 0.073 \times 0.017$ mm$^3$ up to 15.0 GPa at room temperature. Prior to the high pressure experiment, the sample was mounted on a nylon loop with paratone oil and measured at ambient pressure to confirm its crystal structure. The sample was then loaded in the Diacell One20DAC manufactured by Almax-easyLab with 500 μm culet-size anvils. A 250 μm thick stainless steel gasket was pre-indented to 58 μm and a hole of 210 μm was drilled using an electric discharge machining system. The 4:1 methanol–ethanol mixture was used as a pressure transmitting medium. The pressure in the cell was monitored by the R1 fluorescence line of ruby [10,11].

The single crystal XRD measurements were performed using a Rigaku XtalLAB Synergy, Dualflex, Hypix single crystal X-ray diffractometer on Mo K$_α$ radiation ($λ = 0.71073$ Å, micro-focus sealed X-ray tube, 50 kV, 1 mA), operating at $T = 300(1)$ K. The total number of runs and images was based on the strategy calculation from the program CrysAlisPro 1.171.43.104a (Rigaku OD, 2023). Data reduction was performed with correction for Lorentz polarization. For the ambient pressure dataset, numerical absorption correction is based on Gaussian integration over a multifaceted crystal model. Empirical absorption correction using spherical harmonics, implemented in SCALE3 ABSPACK scaling algorithm. The crystal structure was solved and refined using the Bruker SHELXTL Software Package [14,15].

*Density functional calculations*

We performed electronic structure calculations within density functional theory [16,17] (DFT) by using the pseudo-potential augmented plane-wave[18,19] (PAW) basis set from the Vienna Ab initio Simulation Package (VASP) [20-23]. All calculations were performed with the Perdew-Burke-Ernzerhof [23] generalized gradient approximation (GGA), and a plane-wave cutoff of 600 eV was used.

First the bulk structure was relaxed on a $9 \times 9 \times 4$ $k$-point mesh using the conventional unit cell of SrCo$_2$P$_2$. Then a $12 \times 12 \times 6$ $k$-point mesh for the self-consistency field (SCF) was used.

We applied pressure using the PSTRESS option of VASP on a pressure range of 0 to 34 GPa in steps of 2 GPa, where the 0 GPa value was calibrated using the $c/a$ ratio from theory and experiment. In order to account for the difference between the computationally relaxed $c/a$ value, as compared to the experimental one, we shifted the computational c/a curve along the pressure axis until its value matched the experimental one (see Fig. S10). Remarkably, this simple calibration step allows for quantitative agreement between the computationally predicted critical pressure for the ucT-cT transition and the experimentally measured one.

To account for magnetism, we applied the fixed spin moment (FSM) approach. Thereby, we sampled the total energy $E_{tot}$ for a magnetic moment $m$ from 0 to 2 $\mu_B/f.u.$ in steps of 0.1 $\mu_B/f.u.$. To extract information for the magnetic susceptibility $\chi$ we used a fit on the range of 0 to 1 $\mu_B/f.u.$ to apply the Stoner theory [24]. Using the Helmholtz free energy $F = E - TS = E$ in DFT one obtains,

$$\frac{\partial^2 E}{\partial m^2} = \chi^{-1} = \chi_0^{-1} - I \quad (1)$$

where $I$ is the Stoner parameter, $E$ the energy, $m$ the magnetic moment, $\chi$ the magnetic susceptibility, and $\chi_0 = N_\uparrow(E_F)$ the Pauli susceptibility equal to the density of states (DOS) at the Fermi level $E_F$ for the spin up channel. The extended Stoner theory [25,26] can then be applied with

$$E(m) = \frac{1}{2} \int_0^m \frac{m' dm'}{\underline{N}(m')} - \frac{Im^2}{4} \quad (2)$$

where $\underline{N}(m')$ is the density of states (DOS) per spin, averaged between the Fermi level of the spin-up and spin-down subbands.

The pressure stabilized ferromagnetic state can then be found as intersections of the $\underline{N}(m)$ curve with the $1/I$ line, if the slope is negative. It is worth noting that this theory can be considered as a special case of Andersen's force theorem, based on the stationary property of a self-consistent DFT solution.

The Stoner factor $I$ is usually considered to be a transferable intra-atomic parameter, for instance, for 3$d$ metals it is usually 0.8-0.9 eV. For a multicomponent system, however, it is renormalized as $I = \sum_i I_i \left(\frac{N_i}{N}\right)^2$ where $I_i$ is the stoner factor if the atom $I$ and $N_i$ is its partial DOS at the Fermi level. Because hybridization depends on pressure, the net Stoner factor does as well.

Crosschecks with the full potential DFT code WIEN2k [27] where done and have been found to be in agreement with VASP.

## SI Figures

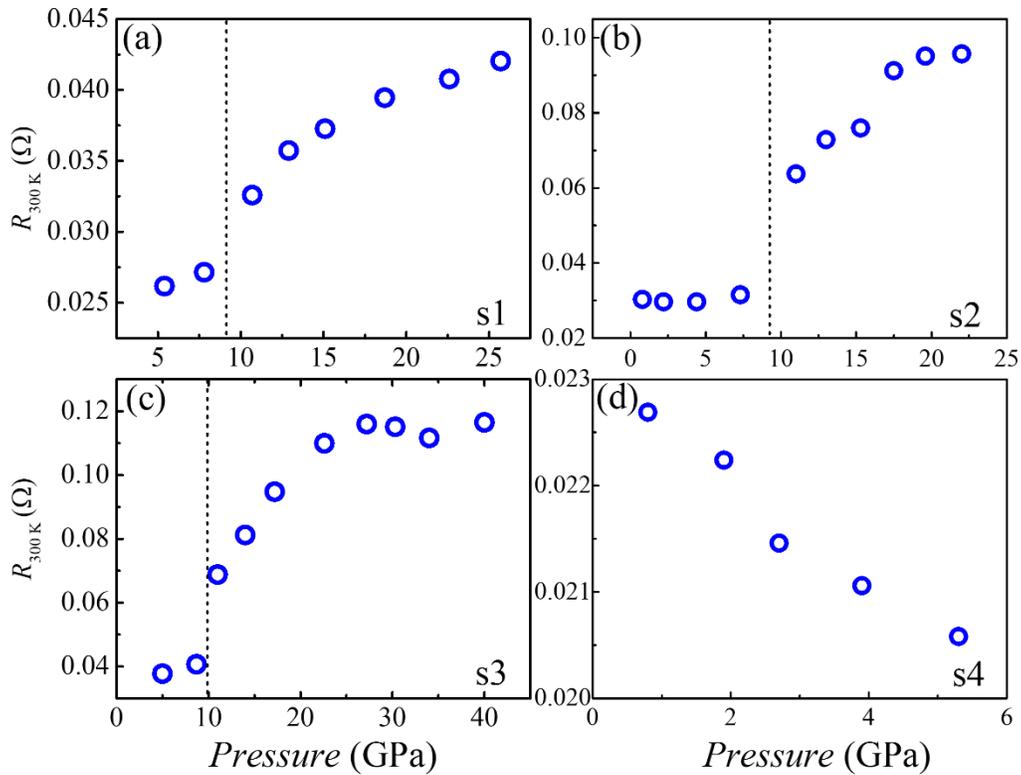

Fig. S1 Pressure dependence of resistance at 300 K, $R_{300K}$ ($p$), on 4 different samples, focusing on different pressure ranges. Detailed: (a) sample s1, diamond culet size 500 μm, (b) sample s2, diamond culet size 500 μm, (c) sample s3, diamond culet size 400 μm, and (d) sample s4, diamond culet size 700 μm. The data were directly collected from the experimental data in Fig. S5.

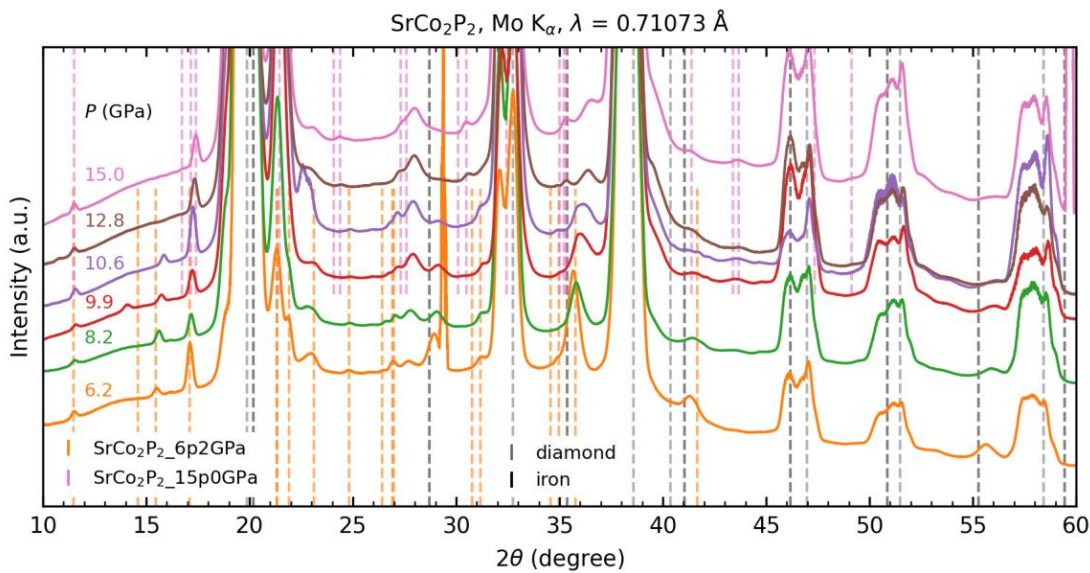

Fig. S2 Pressure dependent powder XRD pattern generated from 2D images obtained in the single crystal XRD measurements.

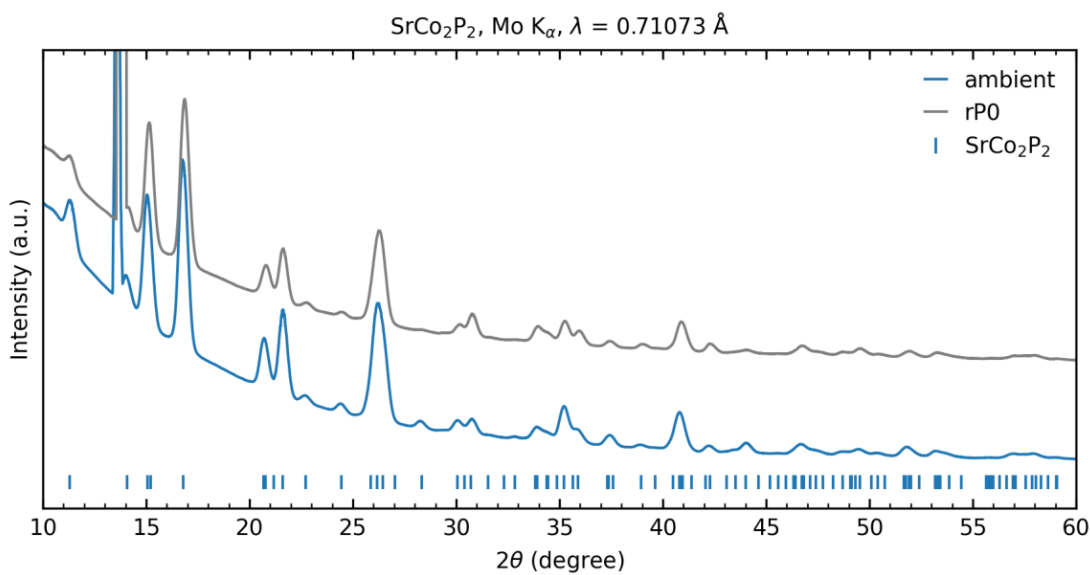

Fig. S3 Powder XRD patterns generated from 2D images obtained in the single crystal XRD measurements at ambient pressure and decompression back ambient pressure after 15 GPa was applied.

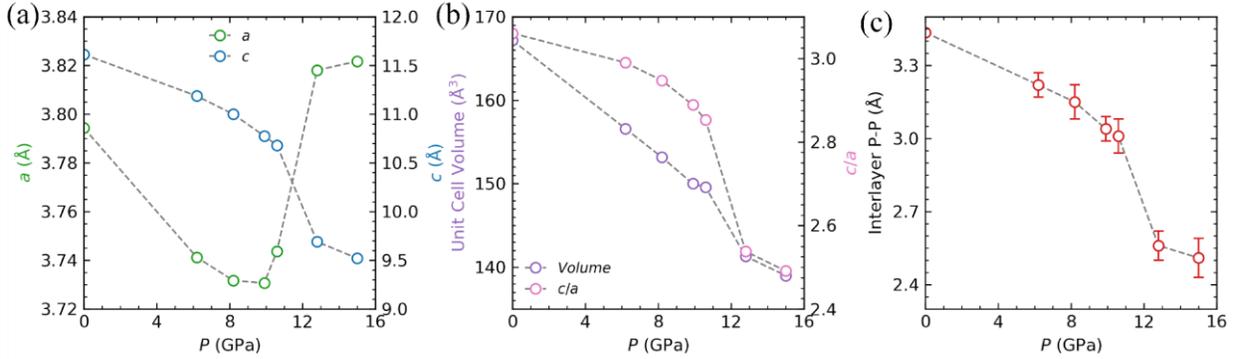

Fig. S4 Pressure dependence of the (a) lattice parameters $a$ and $c$, (b) unit cell volume ($V$) and $c/a$ ratio, and (c) interlayer P-P distance. The results were calculated based on the refinement of the x-ray diffraction data shown in Fig. S2 and S3. The result of $c/a$ ratio as function of pressure is also shown in Fig. 1(a)

We performed single crystal XRD at room temperature under pressure up to 15 GPa. The inferred, powder x-ray diffraction patterns are shown in Figs. S2 and S3. The detailed evolution of lattice parameters, unit cell volume, $c/a$ ratio and P-P distance with the pressure are carefully refined from the diffraction patterns and are shown in Fig. S4. (Tables 1-14 in the SI include the crystallographic data and single-crystal XRD refinement details of the sample at all pressures). We found that $SrCo_2P_2$ maintains the same tetragonal structure (I4/mmm, #139) over the whole measured pressure range. However, a slight expansion of the a-lattice parameter was first observed at 10.6 GPa, accompanied with only a little contraction of the c-lattice parameter. This might be close to the onset of the structure transition to the cT phase. At 12.8 GPa, a dramatic contraction of lattice $c$ (9.4%) and clear expansion of lattice $a$ (1.8%) could be observed, confirming the existence of the ucT to cT phase transition.

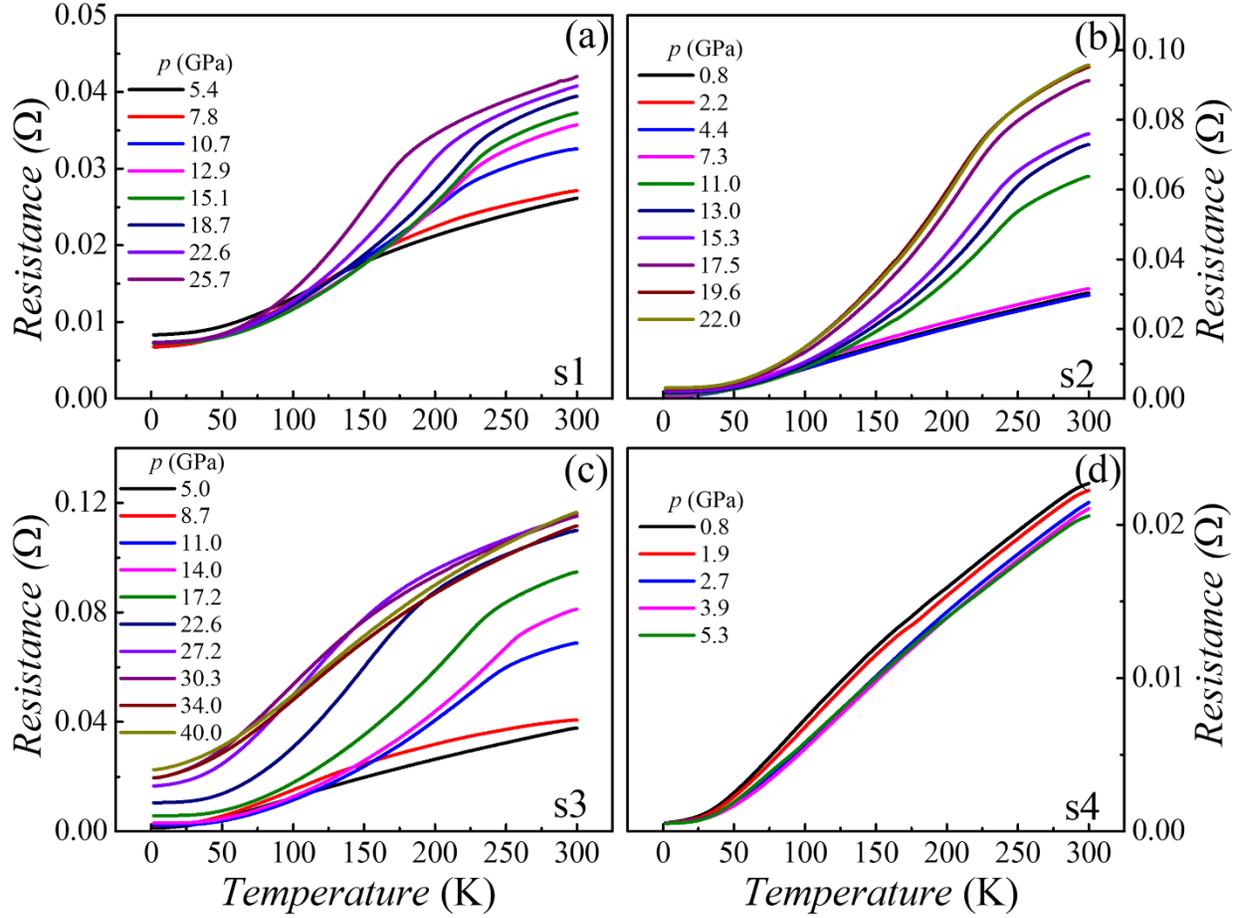

Fig. S5 Temperature dependence of resistance, $R(T)$ under high pressures on 4 different samples, focusing on different pressure ranges. Detailed as (a) sample s1, diamond culet size 500 μm, (b) sample s2, diamond culet size 500 μm, (c) sample s3, diamond culet size 400 μm, and (d) sample s4, diamond culet size 700 μm. $R(T)$ curves in (a) show a relatively larger residual resistance, probably owing to a small bend or exfoliation of the sample.

In Fig. S5, the data of s1 are qualitatively similar to s2 and s3, but there is clear damage and exfoliation during the preparation of the sample before putting contacts. It makes quantitative comparison harder.

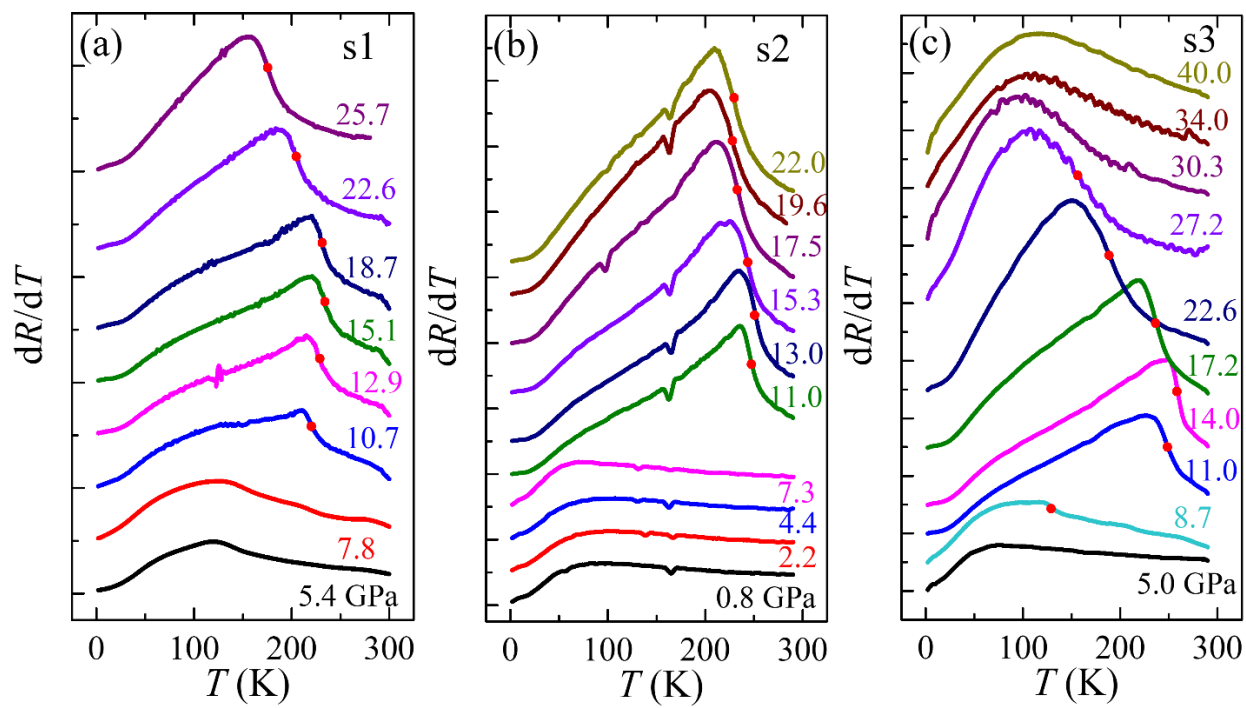

Fig. S6 Temperature dependence of d$R$/d$T$ at various pressures for (a) sample s1, (b) sample s2, and (c) sample s3, based on the $R(T)$ data in Fig. S5 (a)-(c). Red dots marked the evolution of the FM transition temperature, $T_C$, with $T_C$ defined by the midpoints of the d$R$/d$T$ shoulder. All data are shifted for convenience of reading.

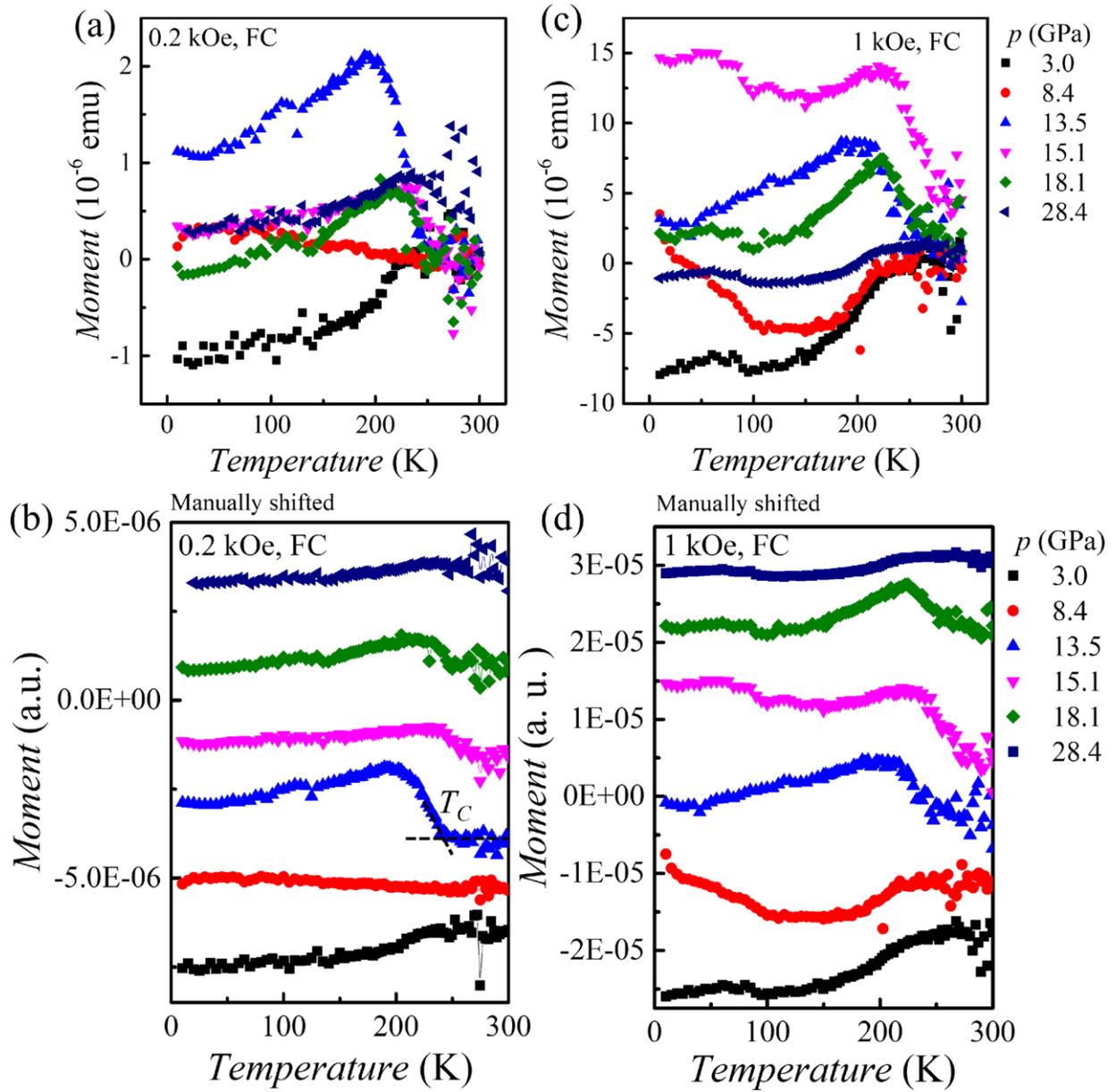

Fig. S7 The field cooling (FC), temperature dependence of magnetic moment, $M(T)$, measured at (a), (b) 0.2 kOe, and (c), (d) 1 kOe, with field direction along [00l] at various pressures on sample s5. The black dashed line in (b) demonstrates the criterion of defining $T_C$ (intersection of two extended lines along $M(T)$ data). (a) and (c) are original data. The curves in both (b) and (d) are manually shifted along the Y-axis for the convenience of clearly seeing the trends and differences.

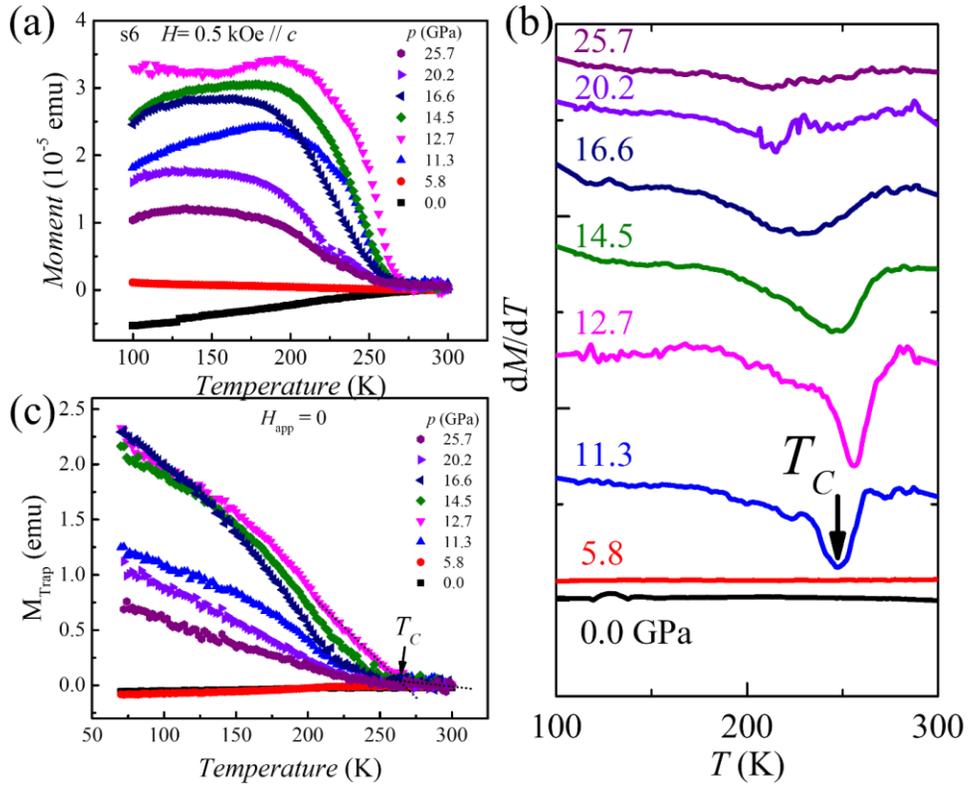

Fig. S8 (a) The field cooling (FC), temperature dependence of magnetic moment, $M(T)$, measured at 0.5 kOe with field direction along [00l] at various pressures on sample s6. The FM transition $T_c$ is defined as the minimum of d$M$/d$T$ valley shown in (b). (c) Temperature dependence of thermoremanent moment, $M_{\text{Trap}}(T)$ at various pressures on sample s6. The thermoremanent magnetization measurement was performed by cooling the DAC from 300 K to 70 K in a 20 kOe external field, stabilizing the temperature at 70 K for 10 minutes, and then removing the field. The warming up $M_{\text{Trap}}(T)$ measurement was carried out from 70 K to 300 K in zero applied field.

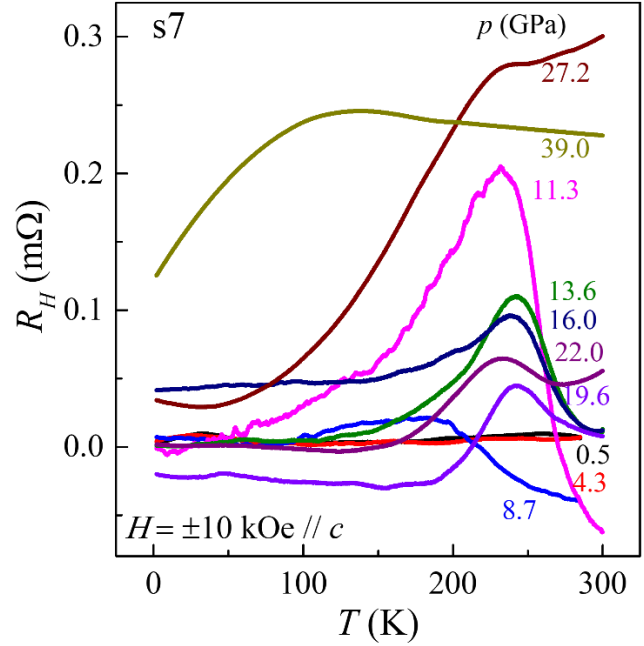

Fig. S9 Temperature dependent Hall coefficient $R_H$ at various pressures up to 39 GPa on sample s7. A ± 10 kOe magnetic field was applied along [00l] direction to measure the $R_H$.

In addition to the magneto-transport data shown in Fig. 3, the temperature-dependent Hall coefficient ($R_H$ ($T$)) under pressure is shown in Fig. S9. At lower pressures (0.5 and 4.3 GPa), the $R_H$, inferred from ± 10 kOe applied field data, is positive indicating the domination of hole carriers, and exhibits a weak temperature dependence in the whole temperature range. At 8.7 GPa, $R_H$ becomes negative at 300 K and shows a transition-like rapid increase upon cooling until ~200 K, below which $R_H$ shows less change. At 11.3 GPa, the $R_H$ upturn becomes much more pronounced. In addition, $R_H(T)$ shows a clear cusp-like feature with peak position at ~ 250 K. The gradually weakening upturn behavior up to 22.0 GPa could be attributed to the anomalous Hall effect resulting from the spontaneous magnetization of the FM order [28], based on the empirical formula:

$$R_H = R_0 H + R_S M \tag{1}$$

where $H$ is magnetic field, $R_S$ is the anomalous Hall coefficient, and $M$ is the magnetization, which is shown in Fig. 2(b) and Fig S7, S8, albeit for much smaller fields. Whereas, the cusp-like behavior, characterized by the rapid decrease in $R_H$ upon cooling below $T_C$, contrasts with typical $R_H(T)$ curves, which are usually flat below $T_C$. This anomaly might be attributed to two different contributions: skew scattering and side-jump mechanisms, which are often considered in the context of the anomalous Hall effect [29]:

$$R_s = a\rho + A\rho^2 \qquad (2)$$

where $\rho$ is resistivity, and $a$, and $A$ are skew scattering and side-jump terms respectively. Further studies are needed to quantitatively understand this. Whereas qualitatively similar behavior was also reported e.g. in ferromagnetic LaCrGe$_3$ [30]. At and above 27.2 GPa, the upturn feature vanishes.

*Density functional theory (DFT) calculations:*

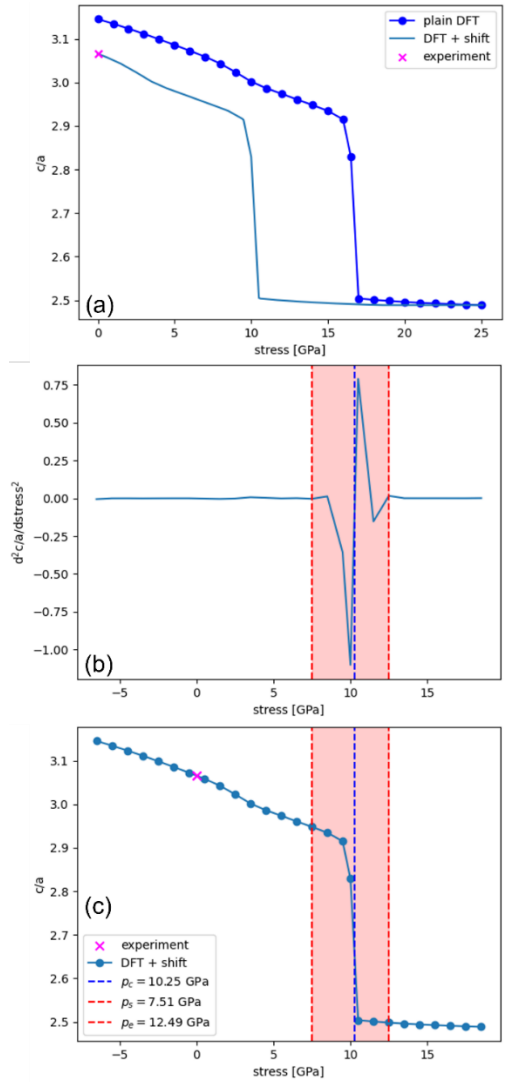

Fig S10: Pressure dependent *c/a* ratio predicted by DFT, with a shift of 6.5 GPa along the x axis to fit the experimental point of *c/a* at ambient pressure. (a) shows the schematic of the pressure calibration. The red region in (b) and (c) marks where the collapse takes place. The dashed lines mark the turning points around the collapse. The blue dashed line $p_c$ is the strongest collapse, while the red line below $p_s$ marks the strongest turning point below $p_c$. This is the same as for $p_e$ but above $p_c$. All $p_s$, $p_c$ and $p_e$ values here are determined using only the shifted DFT data.

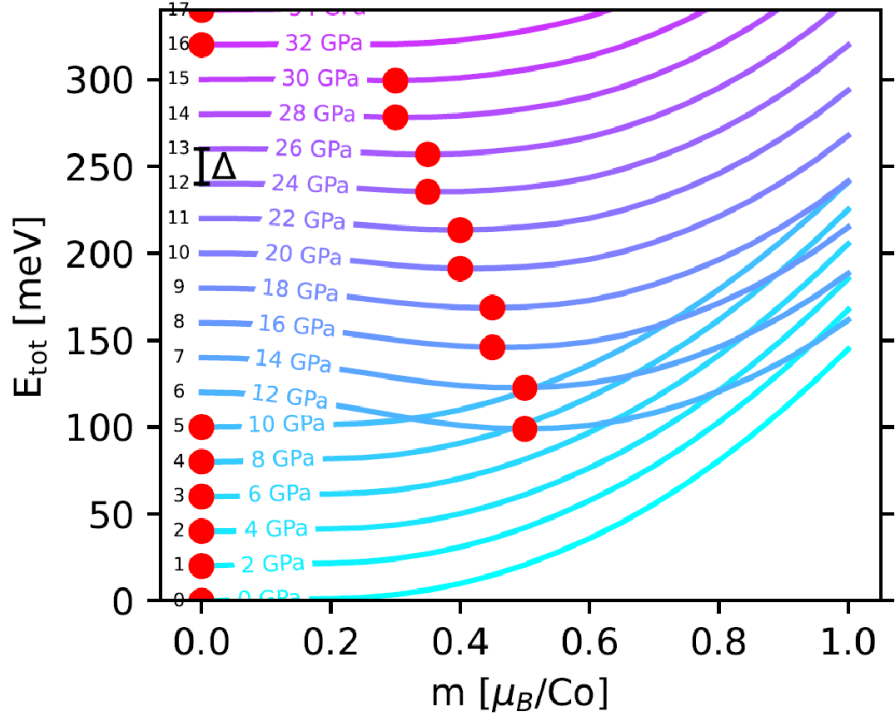

Fig S11: Total energy for fixed spin moment calculations. The x axis is the magnetic moment. The y axis starts the total energy curve starting at the corresponding pressure $p$. The minima in the calculations are marked with a solid, red circle.

**DFT results**

At ambient pressure $SrCo_2P_2$ is in an uncollapsed tetragonal (ucT) phase with large bond distances between the P atoms. After a calibration described in the method section, $SrCo_2P_2$ undergoes a strong volume collapse between 10.2 and 12.2 GPa to a collapsed tetragonal (cT) phase. The electronic structure shows a clear shift in the position of the P $p_z$ bonding and antibonding orbitals, see Fig. 4.

Upon entering the cT phase, $SrCo_2P_2$ develops a long range ferromagnetic (FM) order. The volume collapse of $SrCo_2P_2$ causes a redistribution of the charges, which introduces a self-doping from the P $p_z$ to the Co $d_{xy}$ orbitals. The resulting magnetic moment on Co is about 0.5 $\mu_B$ at the transition (12 GPa) and decreases to about 0.35 $\mu_B$ at 32 GPa. From 34 GPa onward the magnetism vanishes again. The close fitting of the Stoner theory to the DFT data suggests that the FM order is driven by a Stoner instability as discussed below (see Fig. S12).

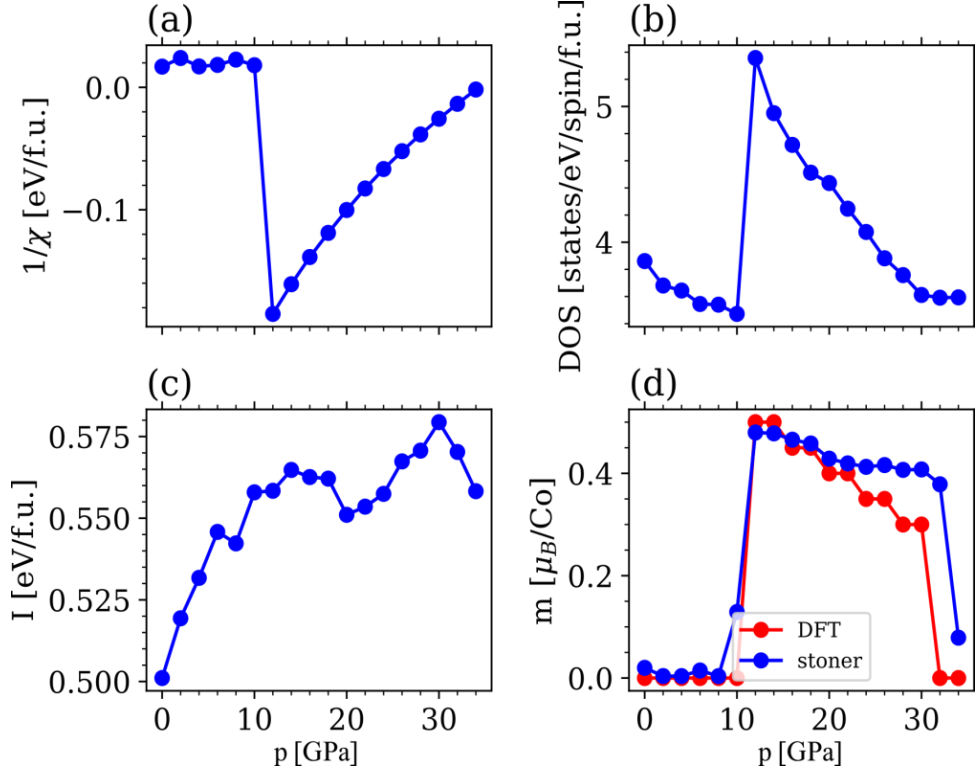

Fig. S12: (a) Magnetic susceptibility $\chi$, plotted as $1/\chi$ obtained from the DFT FSM data, showing a clear transition to a FM state at about 10 to 12 GPa. (b) The average of the density of states (DOS) in a close range ($\pm 60$ meV) around the Fermi level $\underline{N}(E_F)$. (c) Evaluated Stoner parameter $I$ from $1/\chi$ and $\underline{N}(E_F)$. (d) Magnetic moment $m$ obtained from DFT (red) and from Stoner theory (blue) being in good agreement.

The instability causing the FM order to win over the non-magnetic state in DFT is already present at ambient pressure. This is shown by how close to zero the magnetic susceptibility $\chi$ is, see Fig. S12(a). Upon applying pressure, the instability starts to grow, until it reaches its maximum shortly after the collapse to the cT phase. From there on, the instability gets weaker and broadens until it is too weak to maintain the FM long range order.

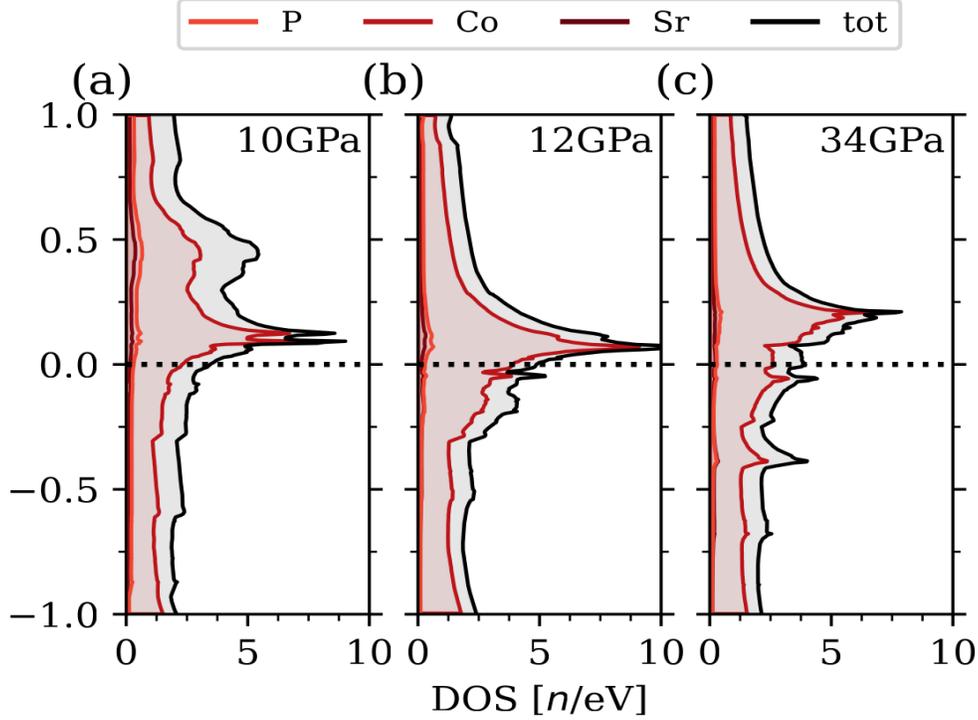

Fig. S13: Total density of states (DOS) and the projected contributions of the atoms Sr, Co and P. Three different DOS are shown. (a) DOS at a pressure just before the collapse at $p_c$. (b) DOS at a pressure $p>p_c$ right after the collapse with the highest instability around the Fermi level. (c) DOS after the brocading out of the instability at the transition to the non-magnetic state in DFT again.

**Interpretation of DFT calculations**

As Fig. S13 shows, there is a strong peak in the DOS right above the Fermi level. This peak originates from the Co $d_{xy}$ orbital. Doping of $\approx 0.5e$ into the system raises the Fermi level into this peak. We will show below that the collapsed phase is self-doped and the peak that used to be entirely unoccupied and not affecting the Stoner instability moves toward the Fermi level and triggers magnetism.

The formal valence of P in this compound is $3^-$, so that the effective Co valence (with the usual reservation of assigning formal valences in a good metal) is $2^+$ [31]. However, in a completely volume collapsed system, as discussed intensively before in the case of $CaFe_2As_2$ [32-36], the pnictogens form covalent dimers, so that the antibonding $p_z$-$p_z$ orbital remains unoccupied and the effective valence of P becomes $2.5^-$ [37]. As a result, the Co bands become self-doped.

**Table 1** The crystal structure and refinement of SrCo$_2$P$_2$ at ambient pressure.

| Chemical Formula | SrCo$_2$P$_2$ |
| --- | --- |
| Formula weight | 267.42 g/mol |
| Space Group | $I4/mmm$ |
| Unit cell dimensions | $a$ = 3.7943(2) Å |
|  | $c$ = 11.6119(10) Å |
| Volume | 167.17(2) Å$^3$ |
| Z | 2 |
| Density (calculated) | 5.313 g/cm$^3$ |
| Absorption coefficient | 26.368 mm$^{-1}$ |
| $F(000)$ | 244 |
| $\theta$ range | 3.51 to 40.50° |
| Reflections collected | 5070 |
| Independent reflections | 193 [$R_{int}$ = 0.0962] |
| Refinement method | Full-matrix least-squares on $F^2$ |
| Data / restraints / parameters | 193 / 0 / 8 |
| Final $R$ indices | $R_1$ ($I>2\sigma(I)$) = 0.0537; $wR_2$ ($I>2\sigma(I)$) = 0.1349 |
|  | $R_1$ (all) = 0.0634; $wR_2$ (all) = 0.1438 |
| Largest diff. peak and hole | +6.824 e/Å$^{-3}$ and -1.048 e/Å$^{-3}$ |
| R.M.S. deviation from mean | 0.660 e/Å$^{-3}$ |
| Goodness-of-fit on F$^2$ | 1.097 |

**Table 2** Atomic coordinates and equivalent isotropic atomic displacement parameters (Å$^2$) of SrCo$_2$P$_2$ at ambient pressure. ($U_{eq}$ is defined as one third of the trace of the orthogonalized $U_{ij}$ tensor.)

|  | **Wyck.** | $x$ | $y$ | $z$ | **Occ.** | $U_{eq}$ |
| --- | --- | --- | --- | --- | --- | --- |
| **Sr** | 2$a$ | 0 | 0 | 0 | 1 | 0.0140(3) |
| **Co** | 4$d$ | 0 | 1/2 | 1/4 | 1 | 0.0122(3) |
| **P** | 4$e$ | 0 | 0 | 0.3521(3) | 1 | 0.0118(5) |

**Table 3** The crystal structure and refinement of $SrCo_2P_2$ at 6.2 GPa.

| Chemical Formula | $SrCo_2P_2$ |
|---|---|
| Formula weight | 267.42 g/mol |
| Space Group | $I4/mmm$ |
| Unit cell dimensions | $a$ = 3.7412(8) Å |
| | $c$ = 11.186(16) Å |
| Volume | 156.6(2) Å$^3$ |
| Z | 2 |
| Density (calculated) | 5.672 g/cm$^3$ |
| Absorption coefficient | 28.153 mm$^{-1}$ |
| $F(000)$ | 244 |
| $\theta$ range | 5.75 to 28.42° |
| Reflections collected | 1078 |
| Independent reflections | 45 [$R_{int}$ = 0.2248] |
| Refinement method | Full-matrix least-squares on $F^2$ |
| Data / restraints / parameters | 45 / 0 / 8 |
| Final $R$ indices | $R_1$ ($I>2\sigma(I)$) = 0.0602; $wR_2$ ($I>2\sigma(I)$) = 0.1708 |
| | $R_1$ (all) = 0.0736; $wR_2$ (all) = 0.1880 |
| Largest diff. peak and hole | +1.389 e/Å$^{-3}$ and -1.088 e/Å$^{-3}$ |
| R.M.S. deviation from mean | 0.310 e/Å$^{-3}$ |
| Goodness-of-fit on F$^2$ | 1.366 |

**Table 4** Atomic coordinates and equivalent isotropic atomic displacement parameters (Å$^2$) of $SrCo_2P_2$ at 6.2 GPa. ($U_{eq}$ is defined as one third of the trace of the orthogonalized $U_{ij}$ tensor.)

| | **Wyck.** | $x$ | $y$ | $z$ | **Occ.** | $U_{eq}$ |
|---|---|---|---|---|---|---|
| **Sr** | 2$a$ | 0 | 0 | 0 | 1 | 0.062(4) |
| **Co** | 4$d$ | 0 | 1/2 | 1/4 | 1 | 0.067(5) |
| **P** | 4$e$ | 0 | 0 | 0.356(2) | 1 | 0.058(8) |

**Table 5** The crystal structure and refinement of SrCo$_2$P$_2$ at 8.2 GPa.

| Chemical Formula | SrCo$_2$P$_2$ |
| --- | --- |
| Formula weight | 267.42 g/mol |
| Space Group | $I4/mmm$ |
| Unit cell dimensions | $a$ = 3.7317(4) Å |
|  | $c$ = 10.999(9) Å |
| Volume | 153.17(14) Å$^3$ |
| Z | 2 |
| Density (calculated) | 5.798 g/cm$^3$ |
| Absorption coefficient | 28.779 mm$^{-1}$ |
| $F(000)$ | 244 |
| $\theta$ range | 5.77 to 27.90° |
| Reflections collected | 548 |
| Independent reflections | 44 [$R_{int}$ = 0.0618] |
| Refinement method | Full-matrix least-squares on $F^2$ |
| Data / restraints / parameters | 44 / 0 / 8 |
| Final $R$ indices | $R_1$ ($I>2\sigma(I)$) = 0.0770; $wR_2$ ($I>2\sigma(I)$) = 0.2013 |
|  | $R_1$ (all) = 0.0964; $wR_2$ (all) = 0.2366 |
| Largest diff. peak and hole | +1.396 e/Å$^{-3}$ and -1.598 e/Å$^{-3}$ |
| R.M.S. deviation from mean | 0.413 e/Å$^{-3}$ |
| Goodness-of-fit on F$^2$ | 1.293 |

**Table 6** Atomic coordinates and equivalent isotropic atomic displacement parameters (Å$^2$) of SrCo$_2$P$_2$ at 8.2 GPa. ($U_{eq}$ is defined as one third of the trace of the orthogonalized $U_{ij}$ tensor.)

|  | Wyck. | x | y | z | Occ. | $U_{eq}$ |
| --- | --- | --- | --- | --- | --- | --- |
| **Sr** | 2a | 0 | 0 | 0 | 1 | 0.063(5) |
| **Co** | 4d | 0 | 1/2 | 1/4 | 1 | 0.070(6) |
| **P** | 4e | 0 | 0 | 0.357(3) | 1 | 0.056(8) |

**Table 7** The crystal structure and refinement of SrCo$_2$P$_2$ at 9.9 GPa.

| Chemical Formula | SrCo$_2$P$_2$ |
| --- | --- |
| Formula weight | 267.42 g/mol |
| Space Group | *I*4/*mmm* |
| Unit cell dimensions | $a$ = 3.7306(7) Å |
| | $c$ = 10.776(16) Å |
| Volume | 150.0(2) Å$^3$ |
| Z | 2 |
| Density (calculated) | 5.922 g/cm$^3$ |
| Absorption coefficient | 29.393 mm$^{-1}$ |
| *F*(000) | 244 |
| $\theta$ range | 5.79 to 30.27° |
| Reflections collected | 1316 |
| Independent reflections | 56 [$R_{int}$ = 0.1645] |
| Refinement method | Full-matrix least-squares on $F^2$ |
| Data / restraints / parameters | 56 / 0 / 8 |
| Final *R* indices | $R_1$ ($I$>2$\sigma$($I$)) = 0.0650; $wR_2$ ($I$>2$\sigma$($I$)) = 0.1470 |
| | $R_1$ (all) = 0.0880; $wR_2$ (all) = 0.1778 |
| Largest diff. peak and hole | +2.747 e/Å$^{-3}$ and -2.288 e/Å$^{-3}$ |
| R.M.S. deviation from mean | 0.465 e/Å$^{-3}$ |
| Goodness-of-fit on F$^2$ | 1.491 |

**Table 8** Atomic coordinates and equivalent isotropic atomic displacement parameters (Å$^2$) of SrCo$_2$P$_2$ at 9.9 GPa. ($U_{eq}$ is defined as one third of the trace of the orthogonalized $U_{ij}$ tensor.)

| | Wyck. | $x$ | $y$ | $z$ | Occ. | $U_{eq}$ |
| --- | --- | --- | --- | --- | --- | --- |
| **Sr** | 2*a* | 0 | 0 | 0 | 1 | 0.058(4) |
| **Co** | 4*d* | 0 | 1/2 | 1/4 | 1 | 0.060(5) |
| **P** | 4*e* | 0 | 0 | 0.359(3) | 1 | 0.060(7) |

**Table 9** The crystal structure and refinement of SrCo$_2$P$_2$ at 10.6 GPa.

| Chemical Formula | SrCo$_2$P$_2$ |
|---|---|
| Formula weight | 267.42 g/mol |
| Space Group | *I*4/*mmm* |
| Unit cell dimensions | a = 3.7437(12) Å |
| | c = 10.68(2) Å |
| Volume | 149.7(3) Å$^3$ |
| Z | 2 |
| Density (calculated) | 5.935 g/cm$^3$ |
| Absorption coefficient | 29.457 mm$^{-1}$ |
| *F*(000) | 244 |
| $\theta$ range | 5.77 to 31.57° |
| Reflections collected | 1364 |
| Independent reflections | 59 [$R_{int}$ = 0.1023] |
| Refinement method | Full-matrix least-squares on $F^2$ |
| Data / restraints / parameters | 59 / 0 / 8 |
| Final *R* indices | $R_1$ (*I*>2$\sigma$(*I*)) = 0.0795; $wR_2$ (*I*>2$\sigma$(*I*)) = 0.2278 |
| | $R_1$ (all) = 0.1033; $wR_2$ (all) = 0.2691 |
| Largest diff. peak and hole | +1.468 e/Å$^{-3}$ and -1.914 e/Å$^{-3}$ |
| R.M.S. deviation from mean | 0.450 e/Å$^{-3}$ |
| Goodness-of-fit on F$^2$ | 1.361 |

**Table 10** Atomic coordinates and equivalent isotropic atomic displacement parameters (Å$^2$) of SrCo$_2$P$_2$ at 10.6 GPa. ($U_{eq}$ is defined as one third of the trace of the orthogonalized $U_{ij}$ tensor.)

| | Wyck. | x | y | z | Occ. | $U_{eq}$ |
|---|---|---|---|---|---|---|
| **Sr** | 2*a* | 0 | 0 | 0 | 1 | 0.067(4) |
| **Co** | 4*d* | 0 | 1/2 | 1/4 | 1 | 0.077(5) |
| **P** | 4*e* | 0 | 0 | 0.359(3) | 1 | 0.069(8) |

**Table 11** The crystal structure and refinement of SrCo$_2$P$_2$ at 12.8 GPa.

| Chemical Formula | SrCo$_2$P$_2$ |
| --- | --- |
| Formula weight | 267.42 g/mol |
| Space Group | $I4/mmm$ |
| Unit cell dimensions | $a$ = 3.8180(10) Å |
| | $c$ = 9.690(17) Å |
| Volume | 141.3(3) Å$^3$ |
| Z | 2 |
| Density (calculated) | 6.287 g/cm$^3$ |
| Absorption coefficient | 31.206 mm$^{-1}$ |
| $F(000)$ | 244 |
| $\theta$ range | 5.74 to 31.77° |
| Reflections collected | 1299 |
| Independent reflections | 54 [$R_{int}$ = 0.0809] |
| Refinement method | Full-matrix least-squares on $F^2$ |
| Data / restraints / parameters | 54 / 0 / 8 |
| Final $R$ indices | $R_1$ ($I>2\sigma(I)$) = 0.0745; $wR_2$ ($I>2\sigma(I)$) = 0.1971 |
| | $R_1$ (all) = 0.0941; $wR_2$ (all) = 0.2484 |
| Largest diff. peak and hole | +1.867 e/Å$^{-3}$ and -1.123 e/Å$^{-3}$ |
| R.M.S. deviation from mean | 0.448 e/Å$^{-3}$ |
| Goodness-of-fit on F$^2$ | 1.430 |

**Table 12** Atomic coordinates and equivalent isotropic atomic displacement parameters (Å$^2$) of SrCo$_2$P$_2$ at 12.8 GPa. ($U_{eq}$ is defined as one third of the trace of the orthogonalized $U_{ij}$ tensor.)

| | Wyck. | $x$ | $y$ | $z$ | Occ. | $U_{eq}$ |
| --- | --- | --- | --- | --- | --- | --- |
| **Sr** | 2$a$ | 0 | 0 | 0 | 1 | 0.055(4) |
| **Co** | 4$d$ | 0 | 1/2 | 1/4 | 1 | 0.064(5) |
| **P** | 4$e$ | 0 | 0 | 0.368(3) | 1 | 0.060(8) |

**Table 13** The crystal structure and refinement of SrCo$_2$P$_2$ at 15.0 GPa.

| Chemical Formula | SrCo$_2$P$_2$ |
| --- | --- |
| Formula weight | 267.42 g/mol |
| Space Group | $I4/mmm$ |
| Unit cell dimensions | $a$ = 3.8217(16) Å |
| | $c$ = 9.52(3) Å |
| Volume | 139.0(4) Å$^3$ |
| Z | 2 |
| Density (calculated) | 6.387 g/cm$^3$ |
| Absorption coefficient | 31.710 mm$^{-1}$ |
| $F(000)$ | 244 |
| $\theta$ range | 5.75 to 30.14° |
| Reflections collected | 1025 |
| Independent reflections | 51 [$R_{int}$ = 0.1988] |
| Refinement method | Full-matrix least-squares on $F^2$ |
| Data / restraints / parameters | 51 / 0 / 8 |
| Final $R$ indices | $R_1$ ($I>2\sigma(I)$) = 0.0746; $wR_2$ ($I>2\sigma(I)$) = 0.1595 |
| | $R_1$ (all) = 0.0924; $wR_2$ (all) = 0.1823 |
| Largest diff. peak and hole | +2.734 e/Å$^{-3}$ and -2.786 e/Å$^{-3}$ |
| R.M.S. deviation from mean | 0.731 e/Å$^{-3}$ |
| Goodness-of-fit on F$^2$ | 1.385 |

**Table 14** Atomic coordinates and equivalent isotropic atomic displacement parameters (Å$^2$) of SrCo$_2$P$_2$ at 15.0 GPa. ($U_{eq}$ is defined as one third of the trace of the orthogonalized $U_{ij}$ tensor.)

| | Wyck. | $x$ | $y$ | $z$ | Occ. | $U_{eq}$ |
| --- | --- | --- | --- | --- | --- | --- |
| **Sr** | 2a | 0 | 0 | 0 | 1 | 0.022(4) |
| **Co** | 4d | 0 | 1/2 | 1/4 | 1 | 0.049(5) |
| **P** | 4e | 0 | 0 | 0.368(4) | 1 | 0.075(11) |